\documentclass[twocolumn,showpacs,preprintnumbers,nofootinbib,superscriptaddress]{revtex4-1}
\usepackage[utf8]{inputenc}
\usepackage[english]{babel}
\usepackage{microtype}	
\usepackage{xspace}
\usepackage{amsmath,amssymb,amsfonts,dsfont,bm} 	% Standard mathematics 
\usepackage{hyperref}
\usepackage{graphicx}% Include figure files
\graphicspath{{figures/}{./}}
\usepackage[normalem]{ulem}
\usepackage{minibox}
\usepackage{color}
\usepackage{xcolor}

\newcommand{\be}{\begin{equation}}  
\newcommand{\ee}{\end{equation}}  
\newcommand{\bear}{\begin{eqnarray}}  
\newcommand{\eear}{\end{eqnarray}}  
\newcommand{\ba}{\begin{array}}  
	\newcommand{\ea}{\end{array}}  
\usepackage{bbold}

\newcommand{\abs}[1]{\left| #1 \right|}

\usepackage{diagbox}

\definecolor{rossoCP3}{cmyk}{0,.88,.77,.40}
\definecolor{blueRef}{rgb}{0.2,0.2,0.6}
\definecolor{blue}{rgb}{0,0.396,0.741}
\hypersetup{
	colorlinks, 
	bookmarksopen, 
	bookmarksnumbered,
	citecolor=blueRef, 		%color of links to bibliography
	linkcolor=rossoCP3,	%color of internal links
	urlcolor=rossoCP3,			%color of external links 
}
\usepackage{siunitx} % <----- Remove this if it does not compile

\newskip\humongous \humongous=0pt plus 1000pt minus 1000pt

\newif\ifdtup

\def\oldreffmt#1{\rlap{[#1]} \hbox to 2\parindent{}}

\def\figfmt#1{\rlap{Figure {#1}} \hbox to 1in{}}  
  
\def\ie{\hbox{\it i.e.}{}\xspace }	\def\etc{\hbox{\it etc.}{}}  
\def\eg{\hbox{\it e.g.}{}\xspace }

\def\Tr{\mathop{\rm Tr}}

\def\abs#1{\left| #1\right|}

\def\beq{\begin{equation}}  
\def\eeq{\end{equation}}  
\def\bea{\begin{eqnarray}}  
\def\eea{\end{eqnarray}}

\def\bq{\begin{quote}}  
\def\eq{\end{quote}}

\def \lta {\mathrel{\vcenter  
     {\hbox{$<$}\nointerlineskip\hbox{$\sim$}}}}  
\def \gta {\mathrel{\vcenter  
     {\hbox{$>$}\nointerlineskip\hbox{$\sim$}}}}   
\relax  
\newcommand{\hc}{\; + \; \mathrm{h.c.} \;}
\newcommand{\LL}{\mathrm{L}}
\newcommand{\RR}{\mathrm{R}}
\newcommand{\U}{\mathrm{U}}
\newcommand{\SU}{\mathrm{SU}}

\newdimen\tdim  
\tdim=\unitlength  
\def\bar{\overline}

\begin{document}

{\title{Where are the Next Higgs Bosons?}

\author{Christopher T. Hill}\email{hill@fnal.gov}
\affiliation{Fermi National Accelerator Laboratory
P.O. Box 500, Batavia, Illinois 60510, USA}
\author{Pedro A. N. Machado}\email{pmachado@fnal.gov}
\affiliation{Fermi National Accelerator Laboratory
P.O. Box 500, Batavia, Illinois 60510, USA}
\author{Anders E. Thomsen}\email{aethomsen@cp3.sdu.dk}
\affiliation{CP$^3$-Origins, University of Southern Denmark
Campusvej 55, DK-5230 Odense M, Denmark
}
\author{Jessica Turner}\email{jturner@fnal.gov}
\affiliation{Fermi National Accelerator Laboratory
P.O. Box 500, Batavia, Illinois 60510, USA}

\date{\today}

\begin{abstract} 
Simple symmetry arguments applied to the third generation
lead to a prediction: 
there exist new sequential Higgs doublets with masses of order
$\lta 5 $ TeV, with approximately universal Higgs-Yukawa coupling constants,
$g\sim 1$. This is calibrated by the known Higgs 
boson mass, the top quark Higgs-Yukawa coupling, and the $b$-quark mass.
A new massive weak-isodoublet, $H_b$, coupled to the $b$-quark
with  $g\sim 1$ is predicted, and 
may be accessible to the LHC at $13$ TeV, and 
definitively at the energy upgraded LHC of $26$ TeV. 
The extension to leptons generates a new $H_\tau$ and
a possible $H_{\nu_\tau}$ doublet. The accessibility of the latter
depends upon whether the mass
of the $\tau$-neutrino is Dirac or Majorana.
\end{abstract}

\preprint{\minibox[]{FERMILAB-PUB-19-143-T \\ CP3-Origins-2019-15 DNRF90}}
\pacs{14.80.Bn,14.80.-j,14.80.-j,14.80.Da}
\maketitle

\section{Introduction}\label{sec:intro}

Understanding flavor physics will likely involve the 
discovery of new particles,  associated with the mystery of the
origin of the small
parameters of the Standard Model. 
For example, 
the observed Higgs-Yukawa coupling of the $b$-quark is small, 
$y_b\simeq 0.024 $.
%\eg,  $(y_b^2/4\pi\alpha)\sim 6.5\times 10^{-3}$. 
This is ``technically natural'' \cite{Hooft} because 
 as $y_b\rightarrow 0 $, a chiral symmetry emerges,  $b_\RR\rightarrow e^{i\theta}b_\RR$,
and a nonzero $y_b$ is never perturbatively regenerated.  It is therefore not 
possible to induce these small parameters as perturbative radiative corrections,
starting from zero, and it becomes challenging
to understand how they are generated.
However, they may
have a perturbative origin, arising from virtual effects involving 
new particles with larger couplings, \ie, a small parameter is induced
by a large parameter that
is power-law suppressed via mixing.
Such new physics may be
accessible to the LHC, or its energy doubled upgrade.

There are, of course, many theoretical ways to achieve this, but 
we are also motivated by the hypothesis that 
a single lone Higgs boson is unlikely to exist---there may be 
a rich spectrum of Higgs bosons, presenting a new spectroscopy in
nature. Thus, the ``new particles'' we will consider 
are exclusively new massive Higgs iso-doublets. 
 
Hence, a plausible origin of the small $y_b$
is via a new heavy Higgs iso-doublet, $H_b$, coupled
as  $ g_b\bar{T}_\LL H_b b_\RR$ where $T_\LL = (t,b)_\LL$. Most importantly, 
the new coupling $g_b$ is large, owing to a symmetry such as we
propose
below, where it is of
order the top quark Higgs-Yukawa coupling,
$g_b= O(1)$.  The observed   $y_b$ is then``effective,'' that is,
$y_b \sim g_b \, (\mu_b^2/M_b^2 )$, 
where the power law suppression
arises from the 
large mass of the new $H_b$, $M_b^2H^\dagger_b H_b$,  and its mixing  
with the Standard Model Higgs doublet $H_0$,
$\mu_b^2 H_0^\dagger H_b + \mathrm{h.c.} $, and $ \mu_b^2/M_b^2 \ll 1 $.
The breaking of the $b_\RR$ chiral symmetry
is then governed, not by $y_b$, but
rather by a large $g_b\sim 1$, yet the small observed 
$y_b\sim 10^{-2}$ arises naturally. 
A UV completion with symmetries and nontrivial renormalization
group evolution has a better chance of predicting the large $g_b$
than the small $y_b$. One would hope
to directly observe the heavy $H_b$ at its mass, $ M_b$.
As a bonus, a larger $g_b$ enhances the production and detection possibilities
for $H_b$ at the LHC, or any other collider.  

Presently we examine this scenario in detail,
which we believe describes the first sequential new 
Higgs doublets that could emerge at the LHC.  
We will focus, for simplicity,
upon the third generation, and will ignore masses and mixings
involving the lighter two generations.
We note that, in a previous study, \cite{HMTT}, the flavor constraints on this system
were found to be consistent.

We assume the top-bottom subsystem is approximately
invariant under a simple extension of the Standard Model symmetry group
structure
\beq
\label{G}
G = \SU(2)_\LL\times \SU(2)_\RR\times \U(1)_{B-L}\times \U(1)_A.
\eeq
By ``approximately''  we mean that, if we  turn off
the $U(1)_Y$ gauging, $g_1\rightarrow 0$,
the symmetry $G$ is exact in the $d=4$ operators
(kinetic terms, Higgs-Yukawa couplings, and potential terms).
The Standard Model (SM) gauging is the usual, $ \SU(2)_\LL \times \U(1)_Y $, and 
is a subgroup of $G$, where
the $\U(1)_Y$ generator is now $I_{3R} + (B-L)/2$.
This electroweak gauging weakly breaks 
the symmetry $ \SU(2)_\RR \times \U(1)_{B-L} \rightarrow \U(1)_Y \times \U(1)_{B-L}$,
however, the $\SU(2)_\RR$ remains as an approximate global symmetry of the $d=4$ operators.
In addition,
a global $\U(1)_A$ arises as well.

To implement $G$ in
the $(t,b)$ sector we require the Standard Model Higgs doublet, $H_0$, couples to $t_\RR$ with coupling  $y_t\equiv g_t$ in the usual way, and a second Higgs doublet, $H_b$,
couples to $b_\RR$ with coupling $g_b$. The symmetry $G$ then dictates 
that there is only a single 
Higgs-Yukawa coupling $g= g_t =g_b$ 
in the quark sector.  This coupling is thus determined by
the known top quark Higgs-Yukawa coupling, $y_t =g \simeq 1$.
A schematic proposal
for a UV completion theory that leads to $G$,
based upon far UV compositeness, was proposed in \cite{HMTT}. In that scheme
 $y_t =g\simeq g_b \simeq 1$ is actually
predicted by the infrared
fixed point  \cite{FP1,FP2}. 
Our current discussion is a simplified subsector of that larger theory.
We do not presently require the ingredients of the larger theory,
and we simply calibrate $g\simeq 1$ from experiment.

Since $m_b/m_t \ll1$,  the $\SU(2)_\RR$ must be broken. 
Here we follow the old rules of chiral dynamics, deploying ``soft'' symmetry breaking
through bosonic mass terms. The  symmetry breaking  in the $d=2$ Higgs mass terms
preserves the universality of quark Higgs-Yukawa couplings $g$.
We remark that these explicit $d=2$ symmetry-breaking mass terms could 
arise from $d=4$ scale invariant interactions involving additional new fields
and hence $G$ could then be broken spontaneously.
However, we are interested presently in the simplest phenomenological scheme
and will be content to insert the $d=2$ symmetry breaking mass terms by hand.

This simple $(t,b)$ quark scenario with  $G$ is then predictive, 
because of the universal coupling.  As we shall observe below, 
the natural mass scale for  $H_b$ is found to be $\sim \SI{5}{TeV}$.  
This prediction involves the assumption that the input Higgs mass, $M_H^2$, is very small,
or  a ``no--fin--tuning'' argument
similar to no--fine--tuning 
of the Higgs mass in the MSSM.

Here we have three {\em a priori} unknown renormalized parameters,
$M_H^2$ (the input Standard Model Higgs potential mass), $\mu_b^2$ (the mixing of $H_b$ with the 
Standard Model Higgs)
and $M_b^2$ (the heavy $H_b$ mass).
We do not specify a UV completion and thus do not
solve the naturalness problems of these quantities
nor explain their origin.  However, once
these are specified, we obtain the $b$-quark mass 
$m_b\sim m_t (\mu_b^2/M_b^2)$, and the
Higgs potential mass, 
$M_0^2=-(\SI{88.4}{GeV} )^2$ $=M_H^2 -\mu_b^4/M_b^2$  (note: $\abs{M_0} =
\SI{125}{GeV} /\sqrt{2}$).
The tachyonic (negative) $M_0^2$ can thus 
be generated from an unknown input value, $M_H^2$, but {\em we should not
fine tune the difference},  $M_H^2 -\mu^4/M^2$.  This requirement
establishes the scale of $M_b \sim \SI{5}{TeV} $.
Or, if we postulate that the input Higgs mass is small, $M_H^2 \ll M_b^2$, we immediately obtain $M_b \simeq \SI{5.5}{TeV}$.

However, this result is modified
when we extend the theory 
to include the third generation leptons $(\nu_\tau, \tau)$.
This requires two additional Higgs fields, $H_\tau$ and $H_{\nu_\tau}$
with masses $M_\tau$ and  $M_\nu$.
The presence of the third generation leptonic Higgs fields with an extension
of the symmetry, $G$,
has the general effect of reducing all the heavy Higgs 
masses, \eg, of $H_b$ to $\sim \SI{3.6}{TeV}$ with $H_\tau$, and
possibly, $H_{\nu_\tau}$, nearby. 

Remarkably, the new Higgs spectrum is then systematically dependent upon whether 
neutrinos receive only Dirac masses, via a ``Dirac-Higgs seesaw mechanism,'' or
Majorana masses in addition to Dirac masses through the Type-I seesaw mechanism.
If neutrino masses are pure Dirac, the  new $H_{\nu_\tau}$ must
be ultra-heavy to produce a seesaw in $\mu^2/M_\nu^2$, and only the
$H_\tau$ is detectable; if on the other hand the neutrino mass
is Majorana, then we expect $M_\nu \sim M_\tau$.
Hence, the physics of
neutrinos and the new Higgs spectrum are
intimately interwoven here.

\section{The top-bottom sub-system}
The assumption of the symmetry, $G$, of Eq.~(\ref{G}) for the top-bottom system
leads to a Higgs-Yukawa (HY)
structure that is reminiscent of  the chiral Lagrangian of the proton and neutron
(or the chiral constituent model of up and down quarks \cite{Manohar})
\bea\label{one}
V_{HY} = g\overline{\Psi}_\LL\Sigma\Psi_\RR \hc\quad
\text{where}
\quad
\Psi= \left( \begin{array}{c}
t \\ 
b%
\end{array}%
\right).
\eea
In Eq.~(\ref{one}), $\Sigma$ is a $2\times 2$ complex matrix and 
$V_{HY}$ is invariant under $\Sigma \rightarrow U_\LL \Sigma U^\dagger_\RR$
with ${\Psi}_\LL \rightarrow U_\LL{\Psi}_\LL$
and ${\Psi}_\RR \rightarrow U_\RR{\Psi}_\RR$.
Note that the $\U(1)_Y$ generator $Y$ of the SM now becomes
$Y= I_{3R} + (B-L)/2 $ and furthermore $\Sigma$ is neutral
under $B-L$ which implies its weak hypercharge $[Y,\Sigma]=\Sigma \, I_{3R}$. 

An additional $ \U(1)_A$ axial symmetry arises as an overall phase transformation
of $\Sigma\rightarrow e^{i\theta}\Sigma$, accompanied
by ${\Psi}_\LL\rightarrow e^{i\theta/2}{\Psi}_\LL$
and ${\Psi}_\RR\rightarrow e^{-i\theta/2}{\Psi}_\RR$.
Keeping or breaking this symmetry is an arbitrary option
for us 
(though  in the $(u,d)$ subsystem this is the Peccei-Quinn symmetry and is required if one incorporates the axion).  

$\Sigma$ can be written
in terms of two column doublets,
\beq
\label{sigma}
\Sigma=(H_0,H^c_b),
\eeq
where  $H^c=i\sigma_2 H^*$. Under $\SU(2)_\LL$ we have 
$H\rightarrow U_\LL H$, and 
$H^c\rightarrow U_\LL H^c$ where we note that the weak hypercharge eigenvalues
of $H$ and $H^c$ have opposite signs.
The HY couplings of Eq.~(\ref{one}) become
\bea
V_{HY} = g_t \left( 
\bar{t}, \, \bar{b}
\right)_{\LL} H_{0} t_{\RR} +g_b \left( 
\bar{t}, \, 
\bar{b}%
\right) _{\LL} H^c_{b} b_{\RR} \hc,
\eea
where the $\SU(2)_\RR$ symmetry has forced $g= g_t= g_b$.

Note that this becomes identical to the Standard Model (SM) 
if we make the identification
\beq
\label{four}
H_b \rightarrow \epsilon H_0,\qquad \epsilon = \frac{m_b}{m_t}=0.0234.
\eeq
The HY coupling of the $ b $-quark to $H_0$ is then $y_b =\epsilon g_b$,
but with $\epsilon \ll 1$ the $\SU(2)_\RR$ symmetry is then lost. 
The SM Higgs boson (SMH), $ H_0 $, in the absence of $H_b$, has the usual SM potential:
\bea
\label{potential0}
V_{\text{Higgs}}= M_{0}^{2}H_0^{\dagger }H_0+\frac{\lambda}{2} (H_0^{\dagger }H_0)^2,
\eea
where $M_{0}^{2}\simeq -( 88.4)$ GeV$^2$ (note the sign), and $\lambda \simeq 0.25 $.
Minimizing this, we find that the Higgs field $H_0$  acquires its usual VEV,  
$v=|M_0|/\sqrt{\lambda}=174$ GeV,  
and the observed physical Higgs boson, $h$, acquires
mass $m_h=\sqrt{2} |M_0| \simeq 125$ GeV.
In our present scheme Eqs.~(\ref{four}) and (\ref{potential0}) arise at low energies dynamically.

Given Eqs.~(\ref{one})and (\ref{sigma}), 
we postulate a new potential of the form
\bea
\label{potential1}
V&=&
M_1^2\Tr\left( \Sigma ^{\dagger }\Sigma \right)
-M_2^2\Tr\left( \Sigma ^{\dagger }\Sigma \sigma_3 \right)
\\ 
&& \!\!\!\!\!\!\!\!\!    \!\!\!\!\!  +\mu ^{2}\left(e^{i\theta} \det\Sigma \hc\right)
+ \frac{\lambda_1 }{2} \Tr\left( \Sigma ^{\dagger }\Sigma\right)^{2}
+ \lambda_2 |\det\Sigma|^2.
\nonumber 
\eea
We have written the potential in the  $\Sigma$ notation  in order
to  display the symmetries more clearly.
In the above, $\sigma_3$
acts on the $\SU(2)_\RR$ side of $\Sigma$, hence
in the limit $M_2^2 = 0$
the potential $V$ is invariant under $SU(2)_R$, while 
 the $\mu^2$ term breaks the 
additional $\U(1)_{A}$. The associated CP-phase
can be removed by field redefinition in the present model.
If the exact $\U(1)_A$ symmetry is imposed
on the  $d=4$ terms in the potential then operators such
as,  $e^{i\alpha} (\det\Sigma)\Tr\left( \Sigma ^{\dagger }\Sigma \right)$,
$e^{i\alpha'} (\det\Sigma)^2$, \etc, are forbidden.
Noting the identity 
$(\Tr( \Sigma^\dagger \Sigma) )^2=\Tr( \Sigma^\dagger \Sigma)^2 + 2\det \Sigma^\dagger \Sigma $,
we obtain only the two indicated $d=4 $ terms as the maximal form of the invariant
potential.

Using Eq.~(\ref{sigma}), $V$ can be written in terms of $H_0$ and $H_b$:
\bea
\label{potb}
V &=& 
M_{H}^{2}H_{0}^{\dagger }H_{0}+M_{b}^{2}H_{b}^{\dagger }H_{b}
+\mu^{2}\left( e^{i\theta} H_{0}^{\dagger }H_{b} \hc \right) 
\nonumber \\
&& +\frac{\lambda }{2}\left(
H_{0}^{\dagger }H_{0}+H_{b}^{\dagger }H_{b}\right) ^{2}+\lambda ^{\prime
}\left( H_{0}^{\dagger }H_{b}H_{b}^{\dagger }H_{0}\right),
\eea
where $\lambda_2=\lambda' + \lambda$, $M_H^2=M_1^2-M_2^2$, and  $M_b^2=M_1^2+M_2^2$.
In the limit $M_b^2\rightarrow \infty$ the field $H_b$  decouples,
and Eq.~(\ref{potb}) reduces to the SM  if
$M_H^2 \rightarrow M_0^2$, and $\lambda \rightarrow 0.25$ thereby recovering Eq.~(\ref{potential0}).

We assume the quartic couplings, following the SM  value $\lambda\sim 0.25$,
contribute negligibly small effects, and set $\theta=0$. Then, varying the potential
with respect to $H_b$, 
the low momentum components of $H_b$ are locked to $H_0$:
%\left\langle H_{b}\right\rangle 
\beq
H_{b}= -\frac{\mu^2}{M_b^2}  H_{0}+ O(\lambda, \lambda').
\eeq
Substituting back into $V$ we recover the SMH potential
\bea
V &=& 
 M_0^2 H_{0}^{\dagger }H_{0}
 +\frac{\lambda }{2}\left(
H_{0}^{\dagger }H_{0}\right)^{2}+O\left(\frac{\mu^2}{M_b^2}\right),
\eea
with
\beq
\label{M0}
M_0^2= M_{H}^{2}-\frac{\mu^4}{M_b^2}.
\eeq
Note that, even with $M_H^2$ positive, $M_0^2=-(88.4)^2$ GeV$^2$
can be driven to its negative value by
the mixing with $H_b$ (level repulsion).
We minimize the SMH potential and define 
\bea
\label{SMHpot}
H_{0}=\left( 
\begin{array}{c}
v +\frac{1}{\sqrt{2}}h\\ 
0
\end{array}%
\right), \qquad v= \SI{174}{GeV},
\eea
in the unitary gauge. 
The minimum of Eq.~(\ref{SMHpot}) 
yields the usual SM result
\beq
v^2 = -M_0^2/\lambda, \qquad m_h  = \sqrt{2}|M_0|=125\;\makebox{GeV},
\eeq
where $m_h$ is the propagating Higgs boson mass.
We can then write
\beq
H_b\rightarrow H_b -\frac{\mu^2}{M_b^2}\left( 
\begin{array}{c}
v +\frac{1}{\sqrt{2}}h\\ 
0
\end{array}%
\right),
\eeq
for the full $H_b$ field. 
This is a  linearized (small angle) approximation to the mixing,
and si reasonably insensitive to the small $\lambda, \lambda'\ll1$.

Note  the effect of ``level repulsion'' of the Higgs mass, $M_0^2$,
downward due to the mixing with heavier $H_b$.
The level repulsion in  the presence of $\mu^2$ and $M_b^2$
occurs due to an approximate ``seesaw'' Higgs mass matrix
\beq
\label{matrix}
\begin{pmatrix}
	M_H^2 & \mu^2 \\
	\mu^2 & M_b^2
\end{pmatrix}.
\eeq
The input value of the mass term $M_H^2$ is
unknown and in principle arbitrary, and can 
have either sign. 
We can presumably bound $M_H^2 \gtrsim \SI{1}{GeV^2}$ from below, 
since QCD effects will mix glueballs and the QCD-$\sigma$-meson with $H_0$ in this limit.
As $M_H^2$ is otherwise arbitrary,
we might then expect that the most probable
value is $M_H^2\sim \SI{1}{GeV^2}$, and $|M_H^2|\ll(\mu^2, M_b^2)$. 

Let
us consider the case  $M_H^2= 0$.
Then  Eq.~(\ref{matrix}) has eigenvalues
$M_0^2=-\mu^4/M_b^2$, and $ M_b^2$.
Thus, in the limit of small, nonzero $|M_H^2|$,
we see that a negative $M_0^2$ arises naturally, and to a good approximation the physical Higgs mass 
is generated entirely by the negative mixing term.

The mass mixing causes the neutral component of $H_b$ to acquire a small VEV (``tadpole'') 
of  $-v(\mu^2/M_b^2)$.  This implies that the SM HY-coupling of the $b$-quark is
induced with the small value $y_b =g_b(\mu^2/M_b^2)$
(note $g_b$ is positive with a phase redefinition of $b_\RR$).
The $b$-quark then receives its mass from $H_{b}$,
\beq
\label{mub1}
m_b =  g_b(m_b)v\frac{\mu^{2}}{M_b^{2}}
=m_{t}\frac{g_b(m_b)\mu^{2}}{g_t(m_t)M_b^{2}},
\eeq
where we have indicated the renormalization group (RG) scales 
at which these couplings should be evaluated.

In a larger framework with a UV completion, such as Ref.~\cite{HMTT},
at a mass scale,  $m>M_b$, both $g_t(m)$ and $g_b(m)$ will have 
a common renormalization group equation modulo $\U(1)_Y$ effects.
We implicitly assume that, at some very high scale $\Lambda \gg v$, 
the $\SU(2)_\LL\times \SU(2)_\RR$ is a good symmetry.
This implies 
$g(\Lambda)= g_t(\Lambda)=g_b(\Lambda) \gta 1$. Then,
we will predict 
$g(M_b)\simeq  g_t(M_b)\simeq g_b(M_b)$, \eg, where the values 
at the mass scale $M_b$ are determined by the RG fixed point \cite{FP1,FP2}
with small splittings due to $\U(1)_Y$.  

Furthermore, we find that $g_t(m)$, and moreso $g_b(m)$, increase somewhat as we evolve 
downward from $M_b$ to $m_t$ or $m_b$; the top quark mass is then $m_t=g_t(m_t) v$ 
where $v$ is the SM Higgs VEV.  From these effects we obtain
the ratio
\beq
\label{Rtau}
R_b = \frac{g_b(m_b)}{g_t(m_t)} \simeq 1.5.
\eeq
The $b$-quark then receives its mass from the tadpole VEV of $H_{b}$.
\beq 
\label{mub}
m_b = g_b(m_b) v\frac{\mu_b^{2}}{M_b^{2}}=m_{t} R_b\frac{\mu_b^{2}}{M_b^{2}}.
\eeq
In the case that the Higgs mass, $M_0^2$,
is due entirely to the level repulsion by $H_b$, \ie $M_H^2 =0$, and
using Eqs.~(\ref{M0}), (\ref{Rtau}) and (\ref{mub}), we obtain
a predicted mass of the $H_b$, 
\beq
M_b = \frac{m_{t}}{m_b} R_b|M_0|  \simeq  \SI{5.5}{TeV},
\eeq
with $m_b = \SI{4.18}{GeV} $, $ m_t=\SI{173}{GeV}$, and $|M_0| = \SI{88.4}{GeV}$.
We remind the reader we have
 ignored the effects of the quartic couplings $\lambda$,
which we expect are small.
Moreover, the quartic couplings do not enter the mixing, because
terms such as, $H^\dagger_0 H_0 H^\dagger_0 H_b $ are forbidden
by our symmetry.  The remaining terms only act as slight
shifts in the masses, never larger than $\sim \lambda v^2$
and can be safely ignored.
%mtau = 1.776==1.78GeV

This is a key prediction of the model
(see also \cite{Porto:2007ed,BenTov:2012cx}).  In fact, we can argue
that with $M_H^2$ nonzero, but with small fine tuning (see below), the result
$M_b\lta \SI{5.5}{TeV}$ is obtained.
This mass scale is accessible to the LHC
with luminosity and energy upgrades, and we feel represents an important
target for discovery of the first sequential Higgs Boson.

However, we will
see in the next section that this result is  reduced
when the third generation leptons are included. 

\section{Inclusion of the $(\nu_\tau, \tau)$ leptons}
The simple $(t,b)$ system described above can be extended to the third generation
leptons $(\nu_\tau, \tau)$. 
Presently we will
abbreviate $\nu_\tau$ to $\nu$, and 
ignore neutrino mixing in this third generation scheme (see also \cite{Porto:2008hb}).
We consider two distinct mechanisms to generate the small neutrino mass scale.

Remarkably the predictions for the mass spectrum
are sensitive to the mechanism of neutrino mass generation.
The next sequential massive Higgs iso-doublet, in addition
to $H_b$, is likely to include the $H_\tau$,
and possibly also $H_\nu$, which is dependent
upon whether neutrino masses are Majorana
or Dirac in nature.   

We introduce the Higgs-Yukawa couplings for the leptons in a
the $G$ invariant form
\bea
\label{oneell}
V_{HY\ell} &=& g_\ell\overline{\Psi}_\LL \Sigma_\ell \Psi_\RR \hc
\nonumber \\
&=& g_{\nu} \left(\bar{\nu}, \, \bar{\tau} \right)_\LL {H}_{\nu} \nu_{\RR}
+ g_{\tau} \left(\bar{\nu}, \, \bar{\tau} \right)_\LL H^c_{\tau} \tau_{\RR} \hc,
\eea
where we've introduced a second ``leptonic'' chiral field 
\beq
\label{sigmaell}
\Sigma_\ell=(H_\nu,H^c_\tau),\quad
\text{where}
\quad
\Psi= \binom{\nu}{\tau},
\eeq
and $\Sigma_\ell$ transforms as $\Sigma$ under $G$.

The couplings $g_{\nu}$ and $g_{\tau}$ are assumed to
have the common value at the high scale $\Lambda$,
\beq
g= g_{t}(\Lambda)=g_{b}(\Lambda)= g_\tau(\Lambda)=g_\nu(\Lambda),
\eeq
Since  the RG equations for the leptons do not
involve QCD, we typically find $ g_\tau(m_\tau) = g_\nu(m_\tau ) \simeq 0.7$ 
for $\Lambda\sim M_\mathrm{Planck}$ \cite{HMTT}.
More generally we  define the parameters
\beq
R_\tau = \frac{g_\tau(m_\tau)}{g_t(m_t)}
\qquad 
R_\nu = \frac{g_\tau(m_\nu)}{g_t(m_t)}
\eeq

We extend the potential of Eq.~(\ref{potential1}),
$V\rightarrow V+ V'$ to incorporate the $\Sigma_\ell$
mass terms: 
\bea
\label{pottau}
V'_0 &=& M_\ell^2 \Tr(\Sigma_\ell^\dagger \Sigma_\ell)
+\delta M_\ell^2 \Tr(\Sigma_\ell^\dagger \Sigma_\ell \sigma_z)
\nonumber \\
&=&
M_{\nu}^{2}H_{\nu}^{\dagger }H_{\nu}+M_{\tau}^{2}H_{\tau}^{\dagger }H_{\tau}
+ ...\;\;
\eea
where the ellipsis refers to quartic terms which  we will
ignore altogether.
Moreover, we assume the new iso-doublets are dormant,
$M_{\nu}^{2}, M_{\tau}^{2} >0$.

We can also introduce a 
mixed term that involves both $\Sigma_\ell$ and $\Sigma$
of Eq.~(\ref{sigma}):
\bea
\label{mu21}
 V_1' &=&
\mu_1^{2}\Tr(e^{i\theta'}\Sigma^\dagger \Sigma_\ell \hc)
\nonumber \\
& =& \mu_1^{2} e^{i\theta'} \left( H_{\nu}^{\dagger } H_{0} +H_{b}^{\dagger } H_{\tau}\right) 
\hc
\eea
Note that this term mixes the Higgs fields as $H_0 \leftrightarrow H_\nu$
and $H_b \leftrightarrow H_\tau$.
Such mixing would lead to 
the $\tau$
acquiring its mass sequentially 
via mixing with $H_b$, which directly mixes with $H_0$
as in Eq.~(\ref{potb}). Although such a scenario
has  potentially interesting physics, we do not pursue
it presently

However, we can introduce a second term consistent with $G$ 
that leads to direct mixing of $H_0^\dagger H_\tau$. This
can be constructed using a charge conjugated $\Sigma^c$, where 
\bea
\Sigma^c_\ell=i\sigma_2\Sigma^*_\ell(-i\sigma_2),
\eea
and note that $\Sigma^c \rightarrow U_\LL\Sigma^c U_\RR$, 
transforms identically to $\Sigma \rightarrow U_\LL\Sigma U_\RR$
under the $\SU(2)$ groups.
The effect of the conjugation is to flip the column and
charge conjugation assigments
of Higgs iso-doublets in $\Sigma_\ell$,
\beq
\label{sigmaellc}
\Sigma^c_\ell =(H_\tau, H^c_\nu).
\eeq
Therefore, a term that permits direct mixing
$H_0 \leftrightarrow H_\tau$
and $H_b \leftrightarrow H_\nu$ is
\bea
\label{mu22}
V_2' &=& \mu_2^2 e^{i\phi'}\Tr(\Sigma^{c\dagger}_\ell \Sigma) \hc
\nonumber \\
& = &
\mu_2^2  e^{i\phi'}(H_{\tau}^{\dagger }H_{0}+H_{b}^{\dagger }H_{\nu})\hc
\eea
It should be noted that this term violates the $\U(1)_A$ symmetry,
but trivially not $\U(1)_{B-L}$ since both $\Sigma's$ are sterile
under $B-L$.  
For simplicity we will simply set the CP phases to zero, 
$\theta=\theta'=\phi'=0$.

\subsection{$H_\tau$, $H_b$, and a Dirac Neutrino Seesaw }

One interesting possibility is that the $\SU(2)_\RR$ symmetry
is badly broken in the lepton sector with 
$M_{\nu}^{2} \gg M_{\tau}^{2}$ of Eq.~(\ref{pottau}). 
In this limit the $H_\nu$ has become ultra-massive and non-detectable
at collider energies.

Through the $ V'_1$ term we observe that $H_\nu$
acquires a tiny tadpole VEV, but now  has  negligible
feedback on the Higgs mass:
\beq
H_\nu= -\frac{\mu_1^2}{M_{\nu}^{2}}H_0, \qquad \;\; \delta M_0^2
= -\frac{\mu_1^4}{M_{\nu}^{2}} \simeq 0.
\eeq
where $\mu_1^2/M^2_\nu\ll1$.
Hence the neutrino acquires a tiny Dirac mass
through the induced coupling to the Higgs:
\bea
V_{HY\ell} &=& -g_\nu \frac{\mu_1^2}{M_{\nu}^{2}}\; \left( 
\begin{array}{c}
\bar{\nu }
\\ \bar{\tau}
\end{array}%
\right)_{\LL} H_{0} \nu_{\RR} +...
\nonumber\\
m_\nu & =& g_\nu v\frac{\mu_1^2}{M_{\nu}^{2}}= m_t R_\nu \frac{\mu_1^2}{M_{\nu}^{2}}.
\eea
wheer $R_\nu \simeq O(1)$.
In such a scenario a  Majorana mass term
is not necessary, as we have a (Dirac) seesaw mechanism
to naturally generate a tiny neutrino mass.

The $H_\tau$  mixes directly 
to the SMH through the $V'_2 $ term.  $H_\tau $ then
acquires a VEV, and feeds back upon the Higgs mass as
\beq
H_\tau= -\frac{\mu_2^2}{M_{\tau}^{2}} H_0, \qquad \;\; \delta M_0^2
= -\frac{\mu_2^4}{M_{\tau}^{2}}. 
\eeq
Hence, the $\tau$ mass is given by
\beq
\label{mutau}
m_\tau =m_{t} R_\tau \frac{\mu_2^{2}}{M_\tau^{2}}.
\eeq
and we expect $R_\tau\sim 0.7$. 

$H_\tau$ and $H_b$ now simultaneously contribute to the SMH mass
\beq
M_{0}^2 =  M_H^2-\frac{\mu_b^{4}}{M_b^{2}}-\frac{\mu_2^{4}}{M_\tau^{2}}.
\eeq
(we remind the reader that $M_0^2=-(88.4)^2$ GeV$^2$ is negative
as defined in Eq.(\ref{potential0})).
Using Eqs.~(\ref{mub1}) and (\ref{mutau}) this yields an elliptical 
constraint on the heavy Higgs masses $M_b$ and $M_\tau$,
\beq
\label{ellipse}
-M_{0}^2 + M_H^2=\frac{m_b^2M_b^{2}}{m_t^2R_b^2}+\frac{m_\tau^2 M_\tau^{2}}{m^2_t R_\tau^2},
\eeq
for fixed $ M_H^2 $. Bear in mind that the Higgs potential input mass, $M_H^2$,
is {\em a priori} unkown,
while $M_0^2= -(\SI{88.4}{GeV})^2 $ is known from the Higgs boson mass, 
$m_h=\sqrt{2}|M_0|=125$ GeV.

If we make the assumption $M_H^2=0$ 
the ellipse is shown as  the red-dashed line in Fig.(\ref{fig:Ft_regions}).
If we further assume both $H_b$ 
and $H_\tau$ contribute equally
to the SMH mass, then we obtain from Eq.~(\ref{ellipse})
\beq
 M_b\simeq \SI{3.6}{TeV}, \qquad   M_\tau \simeq \SI{4.2}{TeV}.
\eeq
Of course, we can raise (lower) these masses by introducing the
bare positive (negative) $M_H^2$.  However, we do not
want to excessively fine tune the difference $M_H^2-\sum_i \mu_i^4/M_i^2  $.

As an alternative way of estimating the Higgs masses we follow the same procedure for estimating the fine tuning
as is sometimes used in the MSSM or composite Higgs models.
To account for the contribution to the fine tuning from all parameters $ x_i $ on an observable $ O $ we use the measure (see e.g. \cite{Barnard:2015ryq})
\begin{equation}
\Delta = \abs{\frac{\partial \ln O}{\partial \ln x_i}}.
\end{equation}
That is to say, the fine tuning is the length of the gradient 
of the logarithmic observable in the space of logarithmic parameters
(other fine tuning definitions have been explored and
lead to quantitatively similar results).

\begin{figure}[t!]
	\includegraphics[width=0.5\textwidth]{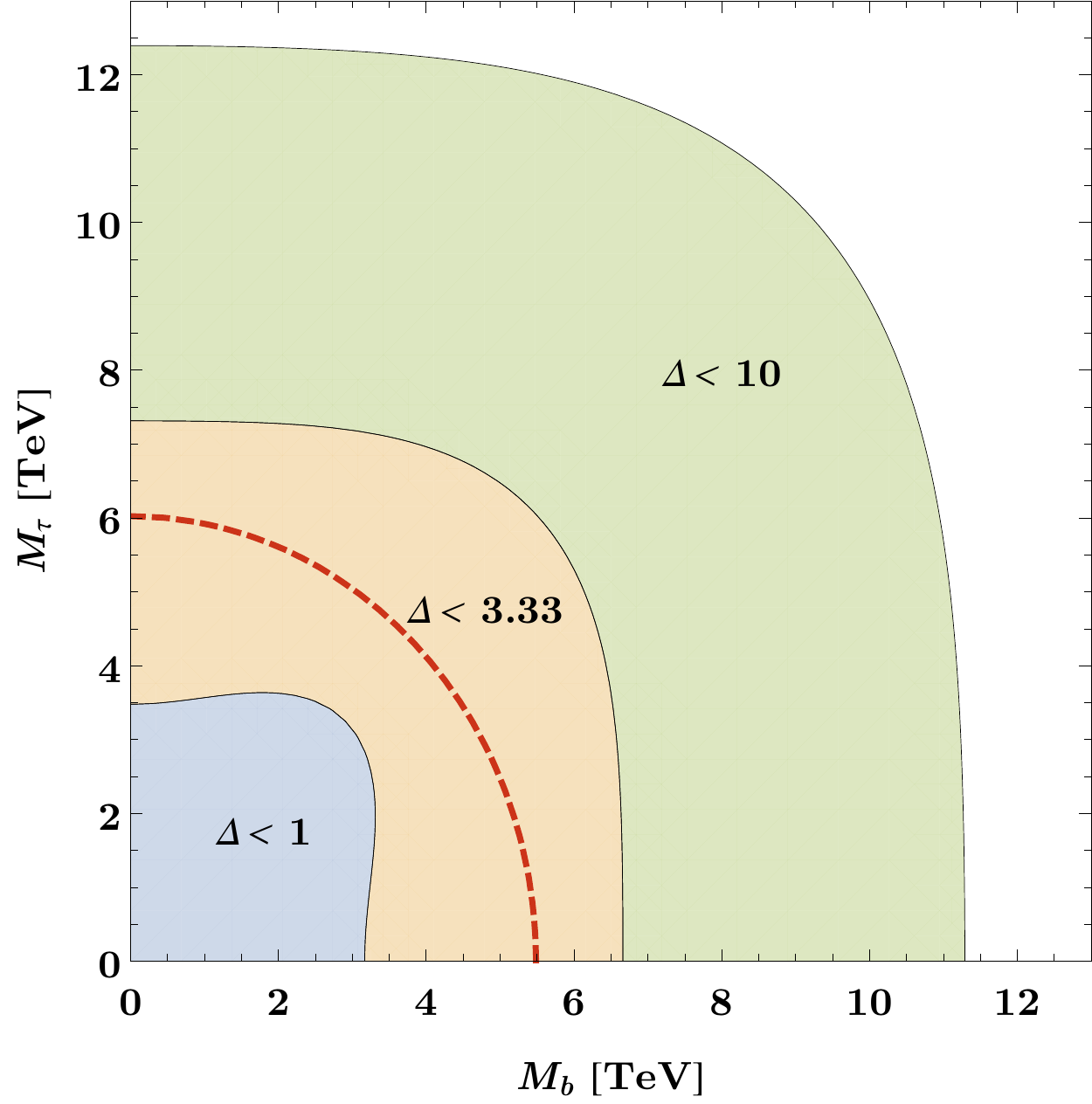}
	\caption{The fine tuning associated with different 
	values of the $ M_b $ and $ M_\tau $ parameters. The 
	remaining three parameters $ \mu_b $, $ \mu_\tau $, 
	and $ M_H $ are fixed by the physical choices of 
	fermion masses and Higgs vev. The red dashed line 
	corresponds to $ M^2_H = 0 $, and the origin to $M_H^2 =-|M_0^2|$.}
	\label{fig:Ft_regions}
\end{figure}
\begin{figure*}[t!]
	\includegraphics[width=1.0\textwidth]{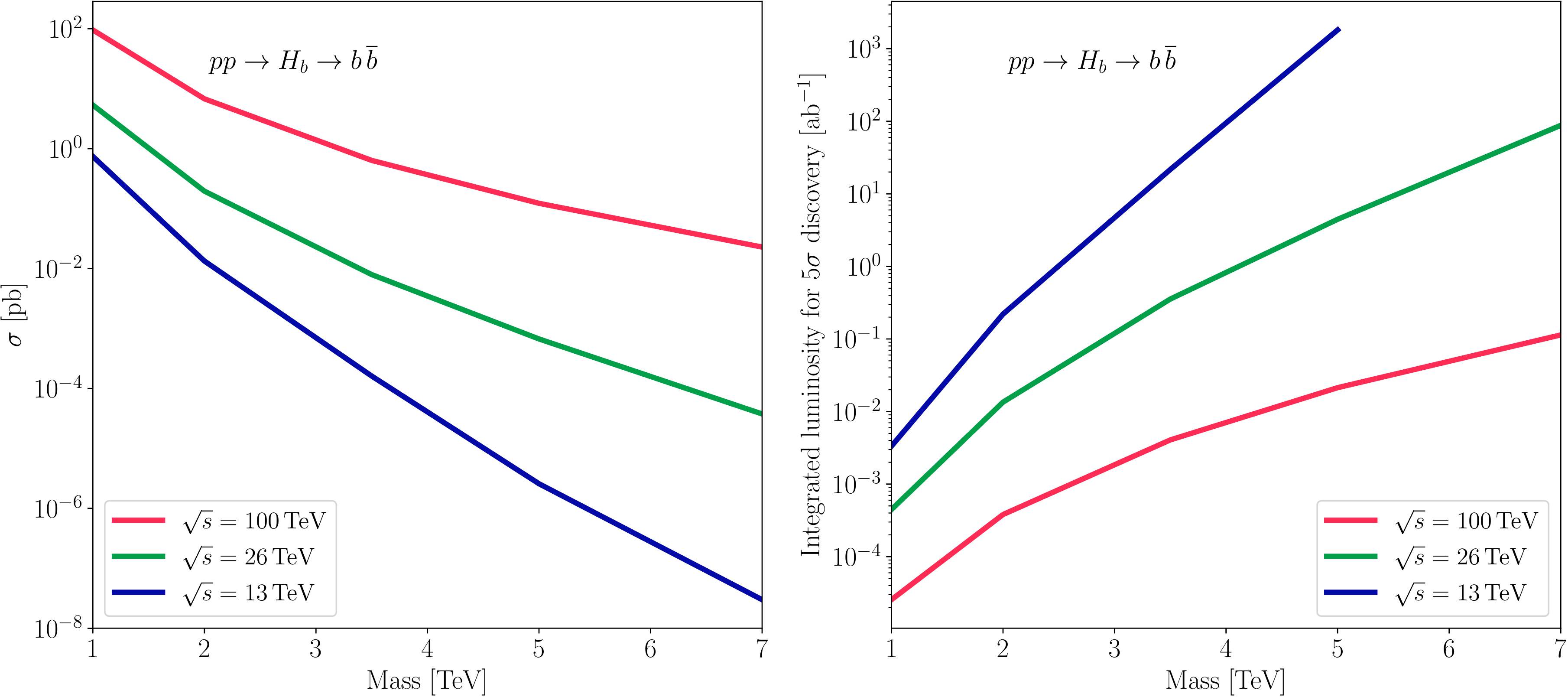}
	\caption{The left plot shows the estimated cross-sections for $pp\rightarrow X + H_b(\rightarrow \bar{b}b)$
	at the LHC $13$ TeV,  LHC upgraded $26$ TeV, and a $100$ TeV $pp$ collider. The 
	cross sections where calculated using structure functions with perturbative $b$ and $\bar{b}$, 
	applying a $100$ GeV $p_T$ cut  on the
	final state $b$-jets, using \texttt{MadGraph5\_aMC@NLO} \cite{Alwall:2014hca} and \texttt{CalcHEP} \cite{Belyaev:2012qa}.
	The right plot displays the estimated $5\sigma$ integrated luminosites for $H_b$ in the $\bar{b}b$
	final state, applying a Breit-Wigner as described in the text, 
	and  assuming a $50\%$ double $b$-tagging efficiency.}
	\label{plot1}
\end{figure*}
Considering the case of  two dormant Higgses, $ H_b $ and $ H_\tau $, 
we measure the log-derivative sensitivities of the known value of $M_0^2$, 
%as a pseudo-observable,
%by way of \eg the $ Z $ mass,
to the five input parameters $M_H^2$, $\mu^2_b$, $\mu_\tau^2$,
$M_b^2$ and $M_\tau^2$. The quadrature combined sensitivities
yield the fine tuning parameter, $\Delta$. 
The fine tuning is summarized in Fig.~(\ref{fig:Ft_regions}) where
 the red-dashed curve indicates $M^2_H = 0$.
Every point on this figure is a bonafide fit to the known values
of $M_0^2$, $m_b$ and $m_\tau$.

Examining just the theory with $ H_b $ alone, we can  
combine the log-derivative sensitivities of $M_0^2$, $m_b$
to three input parameters $M_H^2$, $\mu^2_b$, and
$M_b^2$. Then we find the 
``no fine tuning limit'' 
$ \Delta \leq 1 $ translates into the 
more restrictive constraint 
	\begin{equation}
	 M_b \leq  \SI{3.1}{TeV}.
	\end{equation}	
 The lightest values
of $M_b$ (and $M_\tau$) correspond to $-|M_0^2| = M_H^2$
with no contribution from $M_b^2$  ($M_\tau^2$). 
Indeed, with more heavy Higgses we have generically smaller masses and 
the various become
more easily accessible to the LHC.

\subsection{$H_\nu$, $H_\tau$, $H_b$, and Majorana Neutrinos}
We can also consider the more conventional possibility that
the neutrino has a Majorana mass term, which is an extension of the
physics beyond the minimal model.
The lepton sector Higgs masses may then be comparable,
$M^2_\nu \sim M^2_\tau$, and we have direct coupling to the
SMH by both $H_\nu$ and $H_\tau$, through the
$\mu_1^2$ and $\mu_2^2$ terms. In the absence of a Majorana mass term,  
this would imply $G$ is a good symmetry, with the neutrino
having a large Dirac mass $m_\nu = m_{D\nu}$ comparable to $m_\tau$.

However, we can then suppress the physical
neutrino mass, $m_\nu$, by allowing a the usual Type I seesaw.
We  thus postulate a large Majorana mass term for the
ungauged $\nu_\RR$:
\beq
M\bar{\nu}^c_\RR\nu_\RR \hc
\eeq
Integrating out $\nu_\RR$
we then have an induced $d=5$ operator that
generates a Majorana mass term 
for the left-handed neutrinos \cite{Weinberg:1978kz}
through the VEV of $H_\nu$,
\bea
V_M= \frac{g^2}{2M}\bar{\Psi}_\LL\Sigma_\ell(\mathds{1}-\sigma_3)\Sigma^{c\dagger}_\ell \Psi^c_\LL \hc
\eea

In this scenario there is no restriction that requires the $H_\nu$ mass
to be heavy, and the neutrino physical mass is now small, given by $m_\nu \sim m_{D\nu}^2/M
\sim m_\tau^2/M$
where $m_{D\nu}\sim m_\tau$ is the neutrino Dirac mass. 

Through the $ V'_1$ term we find that $H_\nu$
acquires a  VEV, and has comparable 
feedback on the Higgs mass as $H_\tau$,
\beq
H_\nu= -\frac{\mu_1^2}{M_{\nu}^{2}}H_0,\qquad \;\; \delta M_0^2
= -\frac{\mu_1^4}{M_{\nu}^{2}}.
\eeq
We now have an ellipsoid for the masses
$M_b$, $M_\tau$ and $M_\nu$
\beq
-M_{0}^2 + M_H^2=\frac{m_b^2M_b^{2}}{m_t^2R_b^2}+\frac{m_\tau^2 M_\tau^{2}}{m^2_t R_\tau^2}
+\frac{m_{D \nu}^2 M_\nu^{2}}{m^2_t R_\nu^2}.
\eeq 
With additional light Higgs fields, the elliptical constraint forces all 
of the Higgs masses to smaller values.   We emphasize that
the mass bounds of Fig.~(\ref{fig:Ft_regions}) should
be viewed as upper limits on the Higgs mass spectrum that could
be explored at the LHC.

\section{Phenomenology of the Sequential Higgs Bosons}
Presently we  touch upon the collider phenomenology of this model
and refer the reader to \cite{HMTT} for further discussion.
Flavor physics bounds have also been considered in \cite{HMTT}.

$H_b$ is an iso-doublet with neutral $h^0_b$  and charged $h^\pm_b$
{\em complex field} components
coupling to $(t,b)$ as  $g_b h^0_b \bar{b}_\LL b_\RR + \mathrm{h.c.}$ and $g_b h_b^+ \bar{t}_\LL b_\RR \hc$
with $g_b \simeq 1$.
At the LHC the $h_b^0$ is singly produced in $pp$ by perturbative (intrinsic)  $b$ and $\bar{b}$
components of the proton,
\beq
pp (b\bar{b}) \rightarrow h^0_{b} \rightarrow b\bar{b}.
\eeq
Cross-sections for $13$ TeV, $26$ TeV and $100$ TeV are given for various
$M_b$ in \cite{HMTT} and in Fig.~(\ref{plot1}). 

The cross-section
for production at the LHC of $h_b^0$ is $\sigma(h^0_b\rightarrow b\bar{b})
\sim \SI{e-4}{pb}$ at $13$ TeV  (or $\sigma(h^0_b\rightarrow b\bar{b}) \sim \SI{e-2}{pb}$
at $26$ TeV) for a mass of $M_b ={3.5}$ TeV.
The decay width of $h_b^0$   is large, $\Gamma = 3M/16\pi \simeq \SI{210}{GeV}$, for $M_{b}=3.5$ TeV.
To reduce the background we  
impose a $\SI{100}{GeV}$ $p_T $ cut on the each $b$ jet
in the quoted cross-sections. Note that the charged  $h_b^+ $ would be produced in association with
$\bar{t}b$, has a significantly smaller cross-section
and we have not analyzed it.

We estimate that a $5\sigma$ excess in $S/\sqrt{B}$,
in bins spanning twice the full-width of the Breit-Wigner, requires an
integrated luminosity or $M_b=3.5$ TeV of $\sim \SI{20}{ab^{-1}} $
at $13$ TeV, or
$\sim \SI{100}{fb^{-1}} $ at $26$ TeV.\footnote{ $\SI{1}{ab} = \SI{1}{attobarn} = \SI{e-3}{fb}$ (femtobarn).}  
Similarly, at a mass of $ 5$ TeV one requires 
$\sim \SI{3}{ab^{-1}}$ at $26$ TeV for a $5\sigma$ discovery.
This assumes double $b$-tagging efficiencies of order $50\%$.
Here the background is assumed to be mainly $gg\rightarrow \bar{b}b$, but the large fake
rate from $gg\rightarrow \bar{q}q$ must be reduced to $\lta 3\%$ for this to apply.
 
Note that we have not attempted any
significant optimization of the $S/\sqrt{B}$ in this study.
Hence according to our estimates, while meaningful bounds
are acheiveable at the current $13$ TeV LHC, particularly $M_b\lta 3.5$ TeV,
the energy doubler is certainly favored for this physics and could cover
the full mass range up to  $\sim 7$ TeV. 

Remarkably, the  $h^0_\tau$ (and $h^0_\nu$), neutral 
components of the associated
iso-doublets,   $H_\tau$ and $H_\nu$, may also be singly
produced  because they can mix with 
$H_b$ through the $\mu_1^2$ and $\mu_2^2$ terms of
Eq.~(\ref{mu21}) and (\ref{mu22}) respectively.
This implies a total cross section
\beq
\sigma(h^0_\tau,h^0_\nu) \sim \theta^2 \sigma(h^0_b),
\eeq
for the mixing angle, $\theta$, between either the states $h^0_\tau$ or $h^0_\nu$ and $h^0_b$. 
Although $\theta$ is unknown it could easily be large, $ \theta \sim 0.3$.
The $h^0_\tau\rightarrow \bar{\tau}\tau$ is visible with $\tau$-tagging, and
the background is also slightly suppressed since the peak is narrower by a factor
of $(g_\tau^2/3g_b^2)\sim0.16$.  Hence at the $26$ TeV LHC discovery is
in principle possible for a $3.5$ TeV state with integrated luminosity
of order $2$ ab$^{-1}$.   

 These states are also pair
produced by electroweak processes at the LHC or
through $\gamma + Z^0$ at
a high energy lepton collider:
\bea
\ell^{+}\ell^- &\rightarrow & (\gamma+Z^0)^*\rightarrow H_\tau^0 H_\tau^0\rightarrow
4\tau \nonumber \\
\ell^{+}\ell^- &\rightarrow & (\gamma+Z^0)^*\rightarrow H_x^\pm H_x^\mp, \rightarrow
2\tau +\makebox{missing }p_T 
\nonumber \\
pp &\rightarrow & (W^\pm)^*\rightarrow H_\tau^\pm H_\tau^0\rightarrow 3\tau+\makebox{missing }p_T
\nonumber \\
pp &\rightarrow & (W^\pm)^*\rightarrow H_\nu^\pm H_\nu^0\rightarrow \tau+\makebox{missing }p_T
\eea
where $x$ denotes either $\tau$ or $\nu$.

We note that the LHC now has the capability of ruling out
an $H_b$ with $g_b\sim 1$ of mass $\sim 1$ TeV, with current
integrated luminosities, $\sim \SI{200}{fb^{-1}} $. 

%\newpage

\section{Conclusions}

We believe it is likely that an extended spectrum of Higgs bosons exists in nature.
The expected masses of iso-doublet Higgs bosons have been discussed here with
the prominent new field, $H_b$, couped to $b_R$ in the range $\lta 5$ TeV.
We emphasize that this is an upper limit, and experiments
should search and quote limits in the full mass range for $H_b$.
Additional new states, such as leptonic Higgs iso-doublets, 
or an extended spectroscopy as in \cite{HMTT}, may emerge.

We strongly emphasize that the LHC {\em already has} the capability of ruling out
an $H_b$ with $g_b\sim 1$ of mass $\sim 1$ TeV, with current
integrated luminosities, $\sim \SI{200}{fb^{-1}} $. 
The main reason is that the very plausible $g_b\simeq 1$ significantly
enhances access to these states.  We are unaware of any
limits for these in the literature to date.  We think it is important for
the collaborations to develop an analysis strategy for limiting or discovery of
these states.

This multi-Higgs scenario connects with
the traditional way 
in which physics has evolved from atoms to nuclei to hadrons, 
 thus presenting yet another spectroscopy \cite{Weisskopf}.
Observation of the $H_b$, and measurement of $g_b\simeq 1$,
would constitute compelling
evidence of such an extended Higgs spectrum and validity of
the logic behind these schemes. 

Here
we have shown that a subsector of a more general scalar democracy \cite{HMTT}
can be described simply
in the context of the restricted, yet most observable,
third generation by applying a simple extension of the Standard Model symmetry group.
We think the overall simplicity of this idea is compelling, and we emphasize that
the keystone is the universal Higgs-Yukawa coupling of order unity,
controlled by a symmetry.

Above all, this theory
is testable  and should provide motivation to go further 
and deeper into
the energy frontier with LHC upgrades,  possibly with
a future machine at the $\sim 100$ TeV scale
and/or  high energy lepton colliders.
This scenario also suggests remarkable synergies with the 
ultra-weak scale of neutrinos.
For example, if the $H_\nu$ were inferred 
to exist with a mass 
in the $\lta 10$ TeV range, then
neutrinos must have Majorana masses.

The idea that small effective Higgs-Yukawa couplings
are dynamical, as  $y\sim g(\mu^2/M^2)$, is an old idea inherited from, \eg, 
extended technicolor models \cite{ETC1,ETC2}. The primacy of the top quark
in this scheme, with the calibrating
large Higgs-Yukawa coupling, $g\simeq 1$, was anticipated long ago  
by the top infrared quasi-fixed point \cite{FP1,FP2}.
The fixed point corresponds to a Landau pole in $y_t$ \cite{FP2}
and  the Higgs then naturally arises as a $\bar{t}t$ composite state
\cite{Yama,BHL,HillSimmons}
(\ie, the fixed point \cite{FP2} is the solution to top condensation models)
The model of \cite{HMTT}, which contains the
present model as  a subgroup, has the SMH as a composite  $H^0\sim \bar{T}t$, and 
$H_b\sim \bar{T}b$, $H_\tau\sim \bar{L}\tau$, etc. The fixed point prediction
becomes concordant with the top quark mass in our extended
scheme, and even probes the spectrum of Higgs bosons.

Many of the structural  features of our potentials
are similar to a chiral constituent quark model \cite{Manohar},
though presently $\Sigma$ is subcritical, with only the SM Higgs condensing. 
Indeed, the $\mu^2$ mixing 
term above that generates the $b$-quark mass is equivalent to
a `t Hooft determinant in chiral quark
models, or in topcolor models \cite{TC2}.

Extended Higgs models are numerous, but 
those that most closely presage our present discussion 
are \cite{Porto:2007ed,Porto:2008hb,BenTov:2012cx,Hill1,Rodejohann:2019izm}.
In \cite{HMTT} we suggested a universal composite system of scalars,
which we dubbed ``scalar democracy,'' 
however, in the more minimal third generation scenario presented here,
which is most important for immediate experimental observation, these features are determined
only by the symmetry group $G$. This mainly makes all of the HY couplings
of quarks and leptons universal, modulo RG running, 
and ``hyperfine splitting'' by the small SM gauge interactions. 
We also find it compelling that a negative $M_0^2$ mass can be generated by
this mixing effect starting from a  positive or very small input $M_H^2$.

The universal value of all HY couplings, $g\simeq  1$,
is of central importance to these models,
and also enhances their predictivity.
The observed small parameters of the SM,
such as $y_b$, $y_\tau$, \etc, are given 
essentially by one large universal parameter $g$ multiplied by a suppressing
power law, $\sim \mu^2/M^2$, together 
with smaller perturbative renormalization effects $R_x$. 
In this way, the technically natural small couplings, and
in principle the full CKM structure \cite{HMTT},
can be understood to emerge from the UV and 
the inverted spectrum of Higgs bosons, and may possibly be accessible to experiment.

A large extended spectrum of Higgs bosons with symmetry properties that unify
all HY couplings to large common values is a phenomenologically
rich idea, worthy of further theoretical study and the development
of search strategies at the LHC and other future
high energy colliders.

%\vspace{0.1 in}

%\noindent
\section*{Acknowledgements}

\noindent
This work was done at Fermilab, operated by Fermi Research Alliance,  
LLC under Contract No. DE-AC02-07CH11359 with the United States Department of
Energy.
AET would like to thank Fermilab, 
and gratefully acknowledges 
financial support from the Danish Ministry of Higher Education 
and Science through an EliteForsk Travel Grant. 
The CP$^3$-Origins centre is partially funded by the 
Danish National Research Foundation, grant number DNRF90.

\newpage

\bibliographystyle{apsrev4-1}
\bibliography{bib}{}
\end{document}